\documentclass[aps,amssymb,superscriptaddress,preprint,amsmath]{revtex4}

\usepackage{bm}
\usepackage{graphicx}

\begin{document}
    
\title{Nonanalytic enhancement of the charge transfer from adatom to
one-dimensional semiconductor superlattice and optical absorption spectrum  } 

\author{Satoshi \surname{Tanaka}}
\email{stanaka@p.s.cias.osakafu-u.ac.jp}
\affiliation{Department of Physical Science, Osaka Prefecture 
University, Sakai 599-8531, Japan}
\author{Sterling Garmon}
\affiliation{Center for Studies in Statistical Mechanics and Complex
Systems, The University of Texas at Austin, Austin, TX 78712 USA} 
\author{Tomio Petrosky}
\email{petrosky@physics.utexas.edu}
\affiliation{Center for Studies in Statistical Mechanics and Complex
Systems, The University of Texas at Austin, Austin, TX 78712 USA} 
\affiliation{International Solvey Institute for Physics and Chemistry,
CP231, 1050 Brussels, Belgium}

\date{\today}

\begin{abstract}

The charge transfer from an adatom to a semiconductor substrate of
one-dimensional quantum dot
array is evaluated theoretically.
Due to the Van Hove singularity in the density of electron states at the band
edges, the charge
transfer decay rate is enhanced nonanalytically in terms of the coupling
constant $g$ as $g^{4/3}$.
The optical absorption spectrum for the ionization of a core level 
electron of the adatom to the
conduction band is also calculated.
The reversible non-Markovian process and irreversible Markovian process
in the time evolution of the adatom localized state manifest
themselves in the absorption spectrum through the branch point and pole
contributions, respectively.  

\end{abstract}


\maketitle
\vfill

\section{Introduction}

Thanks to recent advances in nanotechnology, various types of
artificial low-dimensional semiconductor structures have been
fabricated~\cite{Notzel93}.
The quantum confinement of electron in these structures greatly
modifies the density of states of carriers resulting in complete
different electronic and optical properties from the bulk
system~\cite{Ulloa90,Haug92}.
Recently one dimensional quantum wire and quantum-dot array have been
manufactured in various ways~\cite{Notzel93,Gudikesn02},
and the formation of one dimensional miniband has been theoretically
and experimentally
investigated~\cite{Kouwenhoven90,Nikolic98,Joe00}.
It has been found that the Van Hove singularity of the density of state
inherent with the one-dimensionality causes a characteristic electronic
transport~\cite{Hove53,Messica97}.

In this paper, we consider the charge transfer between an adatom localized state
and a one dimensional miniband associated with a quantum wire or
quantum-dot array.
The charge transfer between an adatom and a substrate semiconductor
has been extensively studied~\cite{Miller95}. 
We will show that the low dimensionality greatly modifies
the charge transfer process from the adatom to the semiconductor
quantum dot array due to the singularity of the density of states. 
The physical situation we consider in this paper is shown in
Fig.~\ref{Fig1}(a) where an adatom is attached to a semiconductor
quantum-dot array surface (hereafter we simply call it a superlattice).

The charge transfer of an electron from adatom to the miniband of the
superlattice is caused by the hybridization of the adatom wave function 
with the miniband which can be described by the
bilinear coupling between the adatom localized state and the bound
state of a single quantum dot in which the adatom is situated.
The situation may be described by one-dimensional
version of  
Newns-Anderson model which has
been extensively used to investigate the charge transfer 
process between the adatom and the
semiconductor substrate~\cite{Newns69,Miller95}.
As shown below, the one-dimensional Newns Anderson model we consider
here is equivalent to the Friedrichs Hamiltonian which we
have investigated in our previous letter~\cite{PTG05}.
 
In this recent letter we reported a vast increase in the decay rate 
of an excited dipole molecule traveling in an one-dimensionally confined electromagnetic 
waveguide when the cutoff frequency of the waveguide is near 
the characteristic frequency of the dipole~\cite{PTG05}. 
This vast increase is a direct consequence of a singularity in 
the density of photon states at the cutoff frequency.  
Due to this singularity, standard perturbation analysis 
breaks down and hence one cannot apply Fermi's golden rule to 
evaluate the decay rate in the vicinity of the singularity.
We have shown that in this case the decay rate 
of the excited dipole depends non-analytically as $g^{4/3}$ on the
coupling constant $g$.  
In the present article, we report that the same nonanalytic enhancement of the decay
rate (including the $g^{4/3}$ law) can be found in our system, despite
the fact that the density of electron states of the miniband of the
superlattice is different from the density of states of photon
states in a waveguide.

Although the exponential decay law for the unstable 
state has been observed ubiquitously in nature, quantum mechanics
predicts that there should be a deviation from the exponential
decay law for the unstable state~\cite{Khalfin58}.
It has been shown that, irrespective of any specific form for the
interaction, the time evolution of the surviving amplitude of an 
unstable state deviates from the
exponential law on short and long time scales due to the existence of the
lower bound on the energy, i.e., the branch point effect.
Wilkinson {\it et al.} has recently
 succeeded in measuring the branch point effect in super-cooled
 sodium atoms~\cite{Wilkinson97}, even though the timescale 
 in which the deviation
 from the exponential decay law appears is very short, 
Thus separation of the pole and branch point effects in the
  time evolution of the unstable state is essential to
our understanding of the decay process, as has been done for a system composed of
an excited atom coupled to a radiation field~\cite{Ordonez01,Petrosky01}.
The separation of the two effects is also useful because the
reversible non-Markovian process in the decay process (due to the branch 
point effect) is directly related to the quantum Zeno
effect~\cite{Misra77,Itano90,Petrosky91-2,Koshino05}.

Our goal is to present an actual experimental method that enables 
us to separately evaluate the pole and branch point effects for the
decay process of the unstable state in this system.
In this article, we consider the optical absorption process in which a core
electron of the adatom with discrete energy is transferred to the continuous conduction band.
It will be shown here that the spectral shape is influenced by        
the Markovian process 
due to the pole effect and also by the non-Markovian process due to 
the branch point effect.

The Friedrichs model presented here is equivalent to that known as the
Fano model which was originally developed to explain 
the absorption spectrum for the autoionization process in the He atom~\cite{Fano61}.
It is well known that this absorption spectrum has an asymmetric
spectral profiles  
due to quantum interference
between different optical transition paths.
The appearance of the quantum interference indicates that the quantum
coherence, which is a source of the memory effect, plays a key role in
the decaying process of the excited state.
As will be shown in the present paper, since the branch point effect which
accounts for the non-Markovian decaying process with a memory effect
is intensified by the singularity in the density of states,           
 the absorption spectral profile provides us with the information on
 the extent of the persistence of the 
 quantum coherence in the decaying process.

In {\S 2} we present the model and show the nonanalytic enhancement
of the decay rate for the unstable state.
We decompose the absorption spectrum into contributions from the pole and branch
point effects in {\S 3}.
In {\S 4} we summarize our results and provide some discussion.

\section{Model and nonanalytic enhancement of decay rate}

We consider a 1D semiconductor superlattice with an adatom on the
surface, as shown in Fig.\ref{Fig1}(a).
The width of each potential well is a few nm to 100 nm.  
One can make this device, for example, with GaAs/GaAlAs
heterostructures~\cite{Notzel93,Kouwenhoven90}.
The superlattice consists of $N\gg 1$ identical quantum wells, and 
each well is assumed to have a single bound state $|\tilde{n}\rangle$ 
$(\tilde{n}=  -N$ to $N)$ of equal energy, where the tilde
is used to distinguish the site representation from the wavenumber
representation below.
A miniband is formed in the superlattice due to the electron tunneling
through the potential barrier~\cite{Ulloa90,Nikolic98,Joe00,Zheng02}.
We assume only nearest neighbor tunneling occurs with a transition
probability of $-B/2$.
The 1D superlattice is then represented by the one-dimensional tight
binding model, and we have a continuous miniband of width $2B$        
in the limit of $N\rightarrow\infty$.
In addition to the miniband we consider an adatom localized state 
$|d\rangle$ with energy $E_0$; 
also we consider an inner core level $|c \rangle$ with energy $E_c$.  
Both of these states are associated with the 
adatom impurity located at the $n = 0$ site.  
The adatom localized state $|d\rangle$ is hybridized with 
the $|\tilde{0}\rangle$ state with coupling strength $gB$.


Taking $\hbar =1$ hereafter,
the electronic Hamiltonian $H_E$ is then written as
\begin{eqnarray}\label{Helectron}
    H_E&=&E_c| c\rangle \langle c| + E_0 |d\rangle \langle d|
    -\frac{B}{2}\sum_{<m,n>}|\tilde{m}\rangle \langle \tilde{n}|
    \nonumber \\
    &\quad& +gB\left( |d \rangle \langle \tilde{0} | +| \tilde{0}\rangle
    \langle d | \right) \; ,
\end{eqnarray}
where $<\cdots >$ means taking nearest neighbor sum in Eq.(\ref{Helectron}).
By introducing the wavenumber representation
\begin{equation} \label{k}
    |k\rangle \equiv \frac{1}{\sqrt{N}}\sum_{n=1}^N
    \exp[ikn]|\tilde{n}\rangle  \;,
\end{equation}
we can rewrite $H_E$ in the form of the Friedrichs model
\begin{eqnarray}\label{HE}
    H_E&=&E_c|c\rangle \langle c| + E_d|d\rangle \langle d| +\sum_k E_k 
    |k \rangle \langle k|   \nonumber \\
    &&+\frac{1}{\sqrt{N}}\sum_k gB(|d\rangle
    \langle k| + |k\rangle \langle d| ),
\end{eqnarray}
where $E_k=-B\cos k$.
We impose periodic boundary conditions, leading to $k=2\pi j/N$,
where $j$ is an integer running from $-N/2$ to $N/2$.
The energy dispersion relation gives a divergence in the density of
states at either band edge:
\begin{equation}\label{Rho}
    \rho(E_k)=\frac{1}{\pi}\frac{1}{\sqrt{B^2-E_k^2}}  \; .
\end{equation}


For the 1D Friedrichs model, the electronic Hamiltonian $H_E$ is
diagonalized in terms 
of the so-called Friedrichs solution which results 
in the spectral representation~\cite{Friedrichs48}
\begin{equation}\label{HE2}
    H_E=E_c| c\rangle \langle c| +\sum_{i=1}^2 E_i 
    |\phi_i \rangle \langle \phi_i | +\sum_k E_k
    |\phi_k^F \rangle \langle \phi_k^F |    \;, 
\end{equation}
where $|\phi_i \rangle$ $(i=1,2)$ are two stable eigenstates with 
energies $E_i$, and $|\phi_k^F \rangle$ with energy $E_k$ (where
$|E_k|< B$) is given by
\begin{eqnarray}\label{phikF}
    |\phi_k^F \rangle &=& |k\rangle + \frac{1}{\sqrt{N}}\frac{gB}{\eta^+(E_k)}
        \nonumber\\
	&\times& \left( |d \rangle + \frac{1}{\sqrt{N}}
     {\sum_{k'(\neq k) }} {g B \over{ E_k-E_{k'}+i \epsilon} } | k' \rangle \right) 
    \; ,
\end{eqnarray}
with
\begin{eqnarray}\label{eta}
    &&\frac{1}{\eta^+ (z)}= G^+_{dd}(z)\equiv \langle d|
    \frac{1}{(z-H_E)^+}| d \rangle \nonumber \\
    &&=\frac{1}{\left[z-E_0-\frac{1}{N}\sum_k
	\frac{g^2B^2}{z-E_k}\right]^+}
\end{eqnarray}
and a positive infinitesimal $\epsilon$.
The Green function $G_{dd}^+(z)$ is analytically continued from the
upper half complex $E_k$ plane to the lower half plane.
Here, we do not write explicit forms of $|\phi_i\rangle$, since we
will not use them in this paper~\cite{Friedrichs48}.
In the limit of $N\rightarrow \infty$, the summation over the
wavenumber turns to the continuous integral, and then the self-energy
term in Eq.(\ref{eta}) reads 
\begin{eqnarray}\label{self}
   &&  \Xi(z)\equiv \frac{1}{N}\sum_k
    \frac{g^2B^2}{z-E_k}  \rightarrow  \frac{1}{2\pi}\int_{-\pi}^\pi dk
    \frac{g^2B^2}{z-E_k}  \nonumber \\
 &&    = \int_{-B}^B dE_k \rho(E_k) \frac{g^2B^2}{z-E_k} 
    =\frac{g^2B^2}{\sqrt{z^2-B^2}} \;.
\end{eqnarray}

By substituting the right-hand side of Eq.(\ref{self}) into
$\eta^+(z)$ and taking the square
of the dispersion relation we obtain a quartic equation in $z$.
From the form of the quartic equation, it is readily
seen that the solutions are symmetric about the origin
$E_0 = 0$. 
By applying the standard method for solving
a quartic equation, one can find the explicit solutions of
the dispersion equation $\eta^+(z)=0$.
The solutions consist of the pole $z_d\equiv
\tilde{E}_0-i\gamma$ of
$G^+_{dd}(z)$ in the lower
half complex plane in the second Riemann sheet corresponding to the
unstable decaying state, and the poles on the real axis
corresponding to the stable states, $E_i (i=1,2)$.
For illustration, we plot in Fig.\ref{Fig2}(a) $\tilde{E}_0$ and $E_i$ 
for $g=0.5$ as a function of $E_0/B$
with thick dashed and solid lines, respectively.
For arbitrary $E_0$, there always exist two stable solutions outside the electronic  
miniband.
Note that there is a critical value $E_\gamma$ at which the unstable state  
with imaginary part $\gamma\neq 0$ appears.  
The unstable state exists for all $|E_0| <E_\gamma$, as indicated by 
the arrows in Fig.\ref{Fig2}(a).
 In Fig.~\ref{Fig2}(b) we plot the decay rate $\gamma/B$ of the unstable solution
 for $g=0.1$ as a function of $E_0/B$.
 The maximum value of $\gamma=\gamma_{max}$ occurs at $E_0=\pm B$.

Since the explicit form of the full solution is complicated, we
present an approximate calculation of the maximum value $\gamma_{
max}$ of $\gamma$ at $E_0/B=\pm 1$ and  the critical value $E_\gamma$ of
$E_0$ where the unstable solution appears.   
To estimate $\gamma_{
max}$ we put $E_0= B$ in the dispersion equation.  After a simple
manipulation we obtain
\begin{equation}
\zeta= 1+(-1)^{2/3}{g^{4/3} \over (\zeta+1)^{1/3}},
\label{Itegam}
\end{equation}
where $\zeta\equiv z/B$.  The zeroth order solution (for $g=0$) is
$\zeta=1$.  We use this as our starting point and solve iteratively
to find
\begin{equation}
{\gamma_{ max} \over B}
=   {\sqrt{3}g^{4/3} \over 2^{4/3} } + {g^{8/3} \over 2^{8/3}
\sqrt{3}}  + O(g^{16/3}),
\label{gammax}
\end{equation}
where the third order contribution  ($g^{4j/3}$ with $j=3$)
vanishes.

To estimate $E_\gamma$ near $E_0=B$ we put $z=E_0+g^\alpha z_1$ where
$\alpha >0$ and $z_1$ are unknown variables which are independent of
$g$.  
Then, by  keeping only the predominant contribution to the
dispersion equation, we obtain 
\begin{equation}
\zeta-{\bar E}_0 \approx {g^2  \over \sqrt{{\bar
E}_0+1}\sqrt{\zeta-1}},
\label{apEgam}
\end{equation}
where  ${\bar E}_0\equiv E_0/B$.  Squaring this equation yields
\begin{equation}
f(\zeta)=(\zeta-{\bar E}_0)^2(\zeta-1) -{g^4\over {\bar E}_0+1}
\approx 0. 
\label{apEgam2}
\end{equation}
The function $f(\zeta)$ represents a cubic curve. By taking the
derivative of $f(\zeta)$  and setting $f'(\zeta)=0$, we find the
threshold value of ${\bar E}_0=E_\gamma/B>1$ at which the complex
solutions of  $f(\zeta)=0$ appear.  This leads to  the first two
terms in the following expression for  $E_\gamma$:
\begin{equation}
{E_\gamma\over B} =  1 +   {3g^{4/3} \over 2}   - {g^{8/3}\over
8}+ O(g^{12/3}). \label{Egam}
\end{equation}
For precision, we have presented the second order correction that is
obtained from the exact solution of the original quartic equation.
The values of $\gamma$ and $E_\gamma$ are non-analytic in $g$ for
$g=0$, and hence one cannot obtain these results from a series
expansion in $g$ using  ordinary perturbation analysis. 

 It should be noted that the $g^{4/3}$ law for the decay rate is rather
 universal around the edge of the continuous spectrum.
  Indeed, if the integration over the wavenumber in the self-energy 
  is a
Cauchy integral that is a double-valued function on $z$, 
as is the case in Eq.(\ref{eta}),
then the
edge of the miniband will give a square root type of
essential singularity in the self-energy.  
In addition,  if this singularity leads to a divergence at the edge, 
one can show that the
$g^{4/3}$ effect appears in the vicinity of the singularity.
 Indeed, this is also the case for the dipole molecule traveling
inside the waveguide mentioned above~\cite{PTG05}.

 In terms of the solutions of the dispersion equation obtained here, we
may evaluate the time evolution of the surviving amplitude of the
adatom localized state defined
by $P(t)\equiv\left| \langle d |\exp[-iH_E t] | d\rangle \right|^2$.
Indeed, aside from the persistent oscillation attributed to the two stable
states, there are two contributions to the time evolution of the surviving
amplitude; one
comes from the pole contribution over the wavenumber integral,
where the location of the pole is given by the solution of the
dispersion relation $\eta^+(E_k)=0$ in the complex $E_k$ plane
discussed above, and the other comes from the two branch point
contributions that are located at the edges of the miniband at $\pm
B$.   
As usual, the pole contribution leads to an exponential decay
$\exp[-i(\tilde{E}_0-i\gamma)t]$, while the branch point leads to a
power law decay. 
If the condition $1-|E_0|/B \gg g$ is satisfied, we have a well
separated time scale between the exponential and power law decay 
 in the decay process of the localized state.
 The exponential law for the decay  process  is thus a good approximation
for a  time scale on the order of $t \sim 1/\gamma$.  
However, once this condition is no longer fulfilled, the time separation 
becomes obscure and
we must take into account  the branch point effects in the time
evolution. 
It is, however, a cumbersome task to evaluate the integral associated 
with the branch point in the time evolution. 
In the next section, we will 
evaluate the optical absorption spectrum instead, to separate         
 the contribution of the Markovian effect from the non-Markovian effect. 
 It should be generally easier in experiment to observe the absorption 
 spectrum than a time evolution of the state $| d >$.

\section{Absorption spectrum}
\subsection{Spectral representation and the separation of the pole and
branch point contributions}


Here we consider the optical absorption by a transition from an inner core level of
the adatom to the conduction electronic states of the 1D semiconductor superlattice,
as shown in Fig.\ref{Fig1}(b). 
There are two optical transition paths from the core level $|c\rangle 
$ with large transition
amplitudes: One is
that from the core level $|c\rangle$ to the localized state $|d\rangle$ 
and the other is from
the core level $|c\rangle$ to the $\tilde{n} = 0$ site $|\tilde{0}\rangle$. 
The transition probabilities are denoted by $T_{dc}$ and $T_{0c}$,
respectively.
The contributions of the transitions to the other bound states $|\tilde{n}\rangle$
are small, so we neglect the contributions in this paper, though it is
not difficult to include those effects.

We consider here a single mode of the optical light field with
frequency $\Omega$.
The total Hamiltonian is then written as
\begin{equation} \label{Htotal}
    H_{total}=H_R+H_E+H_{RE}  \;, 
\end{equation}
where the electronic Hamiltonian $H_E$ is given in Eq.(\ref{HE}) and 
\begin{subequations}\label{HR}
    \begin{eqnarray}
	H_R&=&\Omega a^\dagger a  \;, \\
	H_{RE}&=&\left(T_{dc} |d \rangle \langle c | 
	+T_{0c}|\tilde{0} \rangle \langle c | \right) a + \mbox{H.c.} 
	\\
	&\equiv&\hat{T}a+a^\dagger \hat{T}^\dagger \; .
\end{eqnarray}
\end{subequations}
The $H_R$ represents a monochromatic light field where $a (a^\dagger)$ is the
annihilation (creation) operator of the field, and  $H_{RE}$
describes the interaction between
the light field and the electronic system under the rotating wave approximation.

In the initial state of the absorption process, we have a core electron
in the $|c\rangle$ state
and a single photon with energy $\Omega$; we write this initial state as 
$|c;1\rangle$ with
energy $E_c+\Omega$, where $1$ 
denotes the photon number for the mode $\Omega$.
After the absorption process the energy of the photon is transferred
to the electronic system.
The final states then take the form $|\phi_i;0\rangle$ $(i=1,2)$ or $|\phi_k^F;0 \rangle$, 
with the energies $E_i$ or $E_k$ respectively.

When the value of $E_{c}+\Omega$ falls in between $-B$ and $B$, 
the state $|c;1\rangle$ is resonant with the
electronic continuum $|\phi_k^F\rangle$.
As a result, the energy transfer from the light field 
to the electronic system occurs.
Hereafter we take $-E_c$ as the origin of the light energy:
$\Omega+E_c\rightarrow \Omega$.

The decay rate of $|c;1\rangle$ is then determined by 
the pole location of the Green's function for $|c;1\rangle$ as
\begin{equation}\label{Gcc}
    G_{cc}^+(z)\equiv \langle c;1| \frac{1}{[z-H_{total}]^+}|c;1\rangle
    \;.
\end{equation}
The decay rate $\gamma_{abs}(\Omega)$ is a 
function of $\Omega$, and thus we identify the absorption spectrum as
$F(\Omega)\equiv\gamma_{abs}(\Omega)$.
In the weak coupling limit of $T_{dc}$ and $T_{0c}$, the absorption
spectrum $F(\Omega)$ reduces to
\begin{subequations}\label{gammaabs}
    \begin{eqnarray}
	&& F(\Omega)=-\lim_{\epsilon\rightarrow
	0+}\mbox{Im}\sum_k\frac{\left|\langle
	\phi_k^F|\hat{T}|c\rangle\right|^2}{\Omega-E_k+i\epsilon} \\
	&&=-\lim_{\epsilon\rightarrow 0+}\mbox{Im}
	\sum_k\frac{\left| \langle \phi_k^F | \left( T_{dc}  |d\rangle +
	T_{0c}  |\tilde{0} \rangle \right)  \right|^2}
	{\Omega-E_k+i\epsilon}   \; . \quad
    \end{eqnarray}
\end{subequations}
The matrix element in Eqs.(\ref{gammaabs}) may be obtained by
using Eq.(\ref{phikF}):
\begin{subequations}\label{Matrixelement}
\begin{eqnarray}
    \langle d | \phi_k^F \rangle &=& \frac{gB}{\sqrt{N}}
    \frac{1}{\eta^+(z)}  \\
    \langle \tilde{0} | \phi_k^F \rangle &=& \frac{1}{\sqrt{N}} \left[
    1+\frac{\Xi^+(z)}{\eta^+(z)} \right ] \;, \nonumber \\
    &=& \frac{1}{\sqrt{N}} \frac{(E_k-E_0)}{\eta^+(z)} \; ,
\end{eqnarray}
\end{subequations}
where the self-energy $\Xi^+(z)$ is give in Eq.(\ref{self}).

Substituting Eqs.(\ref{Matrixelement}) into Eq.(\ref{gammaabs}b), we
then have
\begin{eqnarray}\label{FOmegaSum}
    &&F(\Omega)=-\mbox{Im} \frac{1}{N}
    \sum_k\frac{1}{\Omega-E_k+i\epsilon} \frac{1}{\eta^+(z)\eta^-(z)} 
    \nonumber \\
    &&\times \left[
    T_{dc}^2+T_{0c}^2\frac{(E_k-E_0)^2}{g^2B^2}
    +2T_{dc}T_{0c}\frac{(E_k-E_0)}{gB}\right]
    \; , \qquad
\end{eqnarray}
where 
$\eta^-(z)$ is analytically continued from the lower half plane.
As done in Eq.(\ref{self}), transforming the summation over the wavenumber 
into the integral in the limit of $N\rightarrow \infty$, the explicit
form of $F(\Omega)$ is obtained as
\begin{eqnarray}\label{FOmega}
    &&F(\Omega)=\frac{g^2B^2\sqrt{B^2-\Omega^2}}
    {(B^2-\Omega^2)(\Omega-E_0)^2+g^4B^4} \nonumber\\
    &&\times \left[
    T_{dc}^2+T_{0c}^2\frac{(\Omega-E_0)^2}{g^2B^2}
    +2T_{dc}T_{0c}\frac{(\Omega-E_0)}{gB}\right]
    \; . \quad
\end{eqnarray}

Now we shall decompose the absorption spectrum $F(\Omega)$
into the pole and the branch point contributions.
For this purpose, we first rewrite $F(\Omega)$ in terms of the contour
integral by using the relation
\begin{eqnarray}\label{sumk}
    && \frac{1}{2\pi}\int_{-\pi}^\pi dk
    \cdots = \int_{-B}^B dE_k \rho(E_k) \cdots  \nonumber\\
    &&=\frac{1}{2\pi i}\int_{-B}^B dE_k
    \frac{1}{g^2B^2}[\eta^+(E_k)-\eta^-(E_k)] \cdots   \;  ,
\end{eqnarray}
where $\cdots$ represents a function of $k$.
By taking the limit of $N\rightarrow \infty$ and applying these relations to 
Eq.(\ref{FOmegaSum}),  $F(\Omega)$ may be cast into the form of a contour integral:
\begin{equation}\label{FOmega2}
    F(\Omega)=\mbox{Im}\frac{1}{2\pi i}\int_\Gamma
    \frac{1}{\Omega-E_k+i\epsilon}\frac{1}{\eta(E_k)} dE_k   \; ,
\end{equation}
where the contour $\Gamma$ is shown in Fig.\ref{Fig3}(a).
As shown in Fig.\ref{Fig3}, the contour $\Gamma$ can be deformed to that shown in 
Fig.\ref{Fig3}(b),
where the cross denotes the pole location at
$E_k=z_d=\tilde{E_0}-i\gamma$.
Along $\Gamma$, the solid and dashed lines are in the first
and second Riemann sheets, respectively, so that $\eta(E_k)$
takes the corresponding value of $\eta^+(E_k)$ and $\eta^-(E_k)$ in
Eq.(\ref{FOmega2}), respectively.

In order to extract the pole contribution from Eq.(\ref{FOmega2}),  
we evaluate the residue around the pole,
 and obtain
\begin{subequations}\label{F0}
    \begin{eqnarray}
	&&F_0(\Omega) \nonumber\\
	&&=\mbox{Im}\frac{1}{2\pi
  i}\int_{pole}dE_k \frac{1}{\Omega-E_k+i\epsilon}\frac{g^2B^2}{\eta^+(E_k)}
 \nonumber \\
 && \times \left[   
 T_{dc}^2+T_{0c}^2\frac{(\Omega-E_0)^2}{g^2B^2}+
 2T_{dc}T_{0c}\frac{(\Omega-E_0)}{gB}\right] \qquad 
    \\
    &&= -\mbox{Im}\frac{N_d}{\Omega-z_d} \left[ T_{dc}^2 
    +T_{0c}^2\frac{(z_d-E_0)^2}{g^2B^2} \right.  \nonumber \\
    && \quad\quad \left. +2T_{dc}T_{0c}\frac{(z_d-E_0)}{gB}
    \right],
\end{eqnarray}
\end{subequations}
where $N_d$ is given by
\begin{subequations}\label{Nd}
\begin{eqnarray}
    N_d^{-1}&=&\frac{d}{dz}\left[ z-E_0-\frac{g^2B^2}{N}\sum_k
    \frac{1}{(z-E_k)^+} \right]_{z=z_d}  \qquad\\
    &=& 1+\frac{z_d (z_d-E_0)}{z_d^2-B^2} \; .
\end{eqnarray}
\end{subequations}
Subtracting the pole contribution from 
Eq.(\ref{FOmega2}), we can write the branch
point contribution as
\begin{equation}\label{F1}
    F_1(\Omega)=\mbox{Im}\frac{1}{2\pi i}\int_{\Gamma'} dE_k
    \frac{1}{\Omega-E_k+i\epsilon}\frac{1}{\eta(E_k)} \;,
\end{equation}
where the contour $\Gamma'$ is depicted in Fig.\ref{Fig3}(c).
The total absorption spectrum is then decomposed to
$F(\Omega)=F_0(\Omega)+F_{1}(\Omega)$.


Though the exponentially decaying unstable state corresponding to the pole in the
second Riemann sheet cannot be identified in the Hilbert space, it is 
possible to identify it outside the Hilbert space.
One of the authors (T.P.) {\it et al.} have shown that the Friedrichs solution can be
decomposed into the unstable state $|\phi_d \rangle$ (and $\langle
\tilde{\phi}_d |$) and the dressed field states $|\phi_k \rangle$ (and $\langle
\tilde{\phi}_k|$)  in the generalized Hilbert space~\cite{Petrosky91}:
\begin{equation}\label{sumphikF}
    \sum_k |\phi_k^F\rangle\langle \phi_k^F |=|\phi_d\rangle \langle
    \tilde{\phi}_d | +\sum_k |\phi_k\rangle \langle
    \tilde{\phi}_k |   \;. 
\end{equation}
One can prove that this decomposition has one to one correspondence
with the $F_0$ and $F_1$ spectral components.
It should be noted that the factor $N_d$ in Eqs.(\ref{Nd}) is now
recognized  
as a normalization constant of the unstable decaying state so that $\langle
\tilde{\phi}_d|\phi_d \rangle =1$, and then $N_d=\langle d |
\tilde{\phi}_d \rangle \langle \phi_d | d \rangle $.  

\subsection{Fano Profile}

In order to clarify the enhancement of the branch point effect due to 
the singularity of the density of states in the 
absorption spectrum, 
we first show in Fig.\ref{Fig4} the calculated results for an
artificial situation with $T_{0c}=0$.
We take $B=1.0$ and $T_{dc}=1.0$ in all calculations in the
present work.
In Fig.\ref{Fig4}, the results for $g=0.2$  
are shown for (a) $E_0=-0.1$ and (b) $E_0=-0.98$.
Solid lines are the total absorption $F(\Omega)$ calculated by
Eq.(\ref{FOmega}), while the dashed and the chain
lines are the pole contribution $F_0(\Omega)$ in 
Eq.(\ref{F0}b) and the branch
point contribution $F_1(\Omega)$ in Eq.(\ref{F1}), respectively.
In Fig.\ref{Fig4}(b), the spectra are magnified around
$\Omega\simeq -B$, while the overall spectrum is shown in the inset.

When $1-|E_0|/B \gg g$ (Fig.\ref{Fig4}(a)), the pole contribution $F_0(\Omega)$ 
is dominant in $F(\Omega)$; the $F_1(\Omega)$
contribution is very small except for the tiny increase around the band
edges.
In this case, $N_d\simeq 1$ and $|\tilde{E}_0-E_0| \ll gB$ 
in Eq.(\ref{F0}b), resulting in a sharp Lorentzian spectrum of $F_0(\Omega)$
as shown in Fig.\ref{Fig4}(a). 
The fact that $F_0(\Omega)$ is dominant in the entire energy region of
$-B \leq \Omega \leq B$, suggests that the time evolution of the adatom
localized state almost completely obeys the exponential decay law, i.e. 
the Markovian
process is predominant, as mentioned in the end of the previous section.

As $E_0$ gets close to the band edge (Fig.\ref{Fig4}(b)), the decay
rate of the unstable state is nonanalytically enhanced (as discussed in \S 2) and the 
energy shift of $|\tilde{E}_0-E_0|$ becomes large due to the singularity
in the density of states of the miniband at the band edges, as
shown in Fig.~\ref{Fig2}.
Therefore the spectral width of $F_0(\Omega)$ gets wider and the
shift of the peak position 
 from $E_0$ becomes visible as shown in Fig.\ref{Fig4}(b). 
Furthermore, it is found from Eq.(\ref{Nd}b) that the divergence in the
density of states enhances the second term in
Eq.(\ref{Nd}b), and thus $N_d$ is largely reduced from 1 while
$\mbox{Im}[N_d]$ remains negligibly small. 
Consequently, the relative ratio of the branch point
contribution of $F_1(\Omega)$ to the pole contribution $F_0(\Omega)$ 
becomes larger than in Fig.\ref{Fig4}(a).
As discussed at the end of the previous section, the relative increase of the branch point 
contribution indicates
that  the time separation between the exponential and the power law
decays becomes obscure in the time evolution of the surviving
amplitude. 
Therefore, the       
non-Markovian process with memory effect becomes significant, which
maintains 
the quantum coherence in the decaying process.
The reduction of $N_d$ ($=\langle d |                  
\tilde{\phi}_d \rangle \langle \phi_d | d \rangle $) also indicates that the unstable state $|\phi_d
\rangle$ (or $\langle \tilde{\phi}_d |$) contains a larger contribution from the electronic continuum
components of $|k\rangle $.
This suggests that the contribution of the continuum to the dressing cloud is more significant.
As shown below, the persistence of quantum coherence in the
decay process, or the large dressing effect, manifests the quantum interference between the
different optical transition paths once the other absorption transition
$T_{0c}$ is introduced.


Next we consider the actual case in which both optical transition paths are
allowed: $T_{0c}\neq 0$ and $T_{dc}\neq 0$.
We show in Fig.\ref{Fig5}(a) the calculated results of $F(\Omega)$
for the same parameters as in Fig.\ref{Fig4}(a) except that $T_{0c}=1.0$ here.
The pole contribution $F_0(\Omega)$ and the branch point contribution 
$F_1(\Omega)$ are shown in Fig.\ref{Fig5}(b) and (c), respectively.
All these spectra are depicted by the black solid lines.
In each panel, the spectra are further decomposed into the spectral components
due to the first, second, and the third terms in $[ \cdots ]$ of
Eq.(\ref{FOmega}) or Eq.(\ref{F0}b).
These terms are attributed to the $d$-$d$ diagonal component,
$\tilde{0}$-$\tilde{0}$ diagonal component, and the interference term between 
these two in
Eq.(\ref{gammaabs}b).  These are shown by the red, blue, and green lines, respectively.

Introduction of $T_{0c}$ changes the symmetric
Lorentzian spectral shape of $F(\Omega)$ shown in Fig.\ref{Fig4}(a) 
, and yields an asymmetric spectral shape around
$\Omega\simeq E_0$ shown by the black line in Fig.\ref{Fig5}(a).
As seen from Fig.\ref{Fig5}(a), this is caused by the interference
term of Eq.(\ref{FOmega}) shown by the green line in the figure.
Around $\Omega\simeq \pm B$ the $F(\Omega)$ shows a sharp rise
which reflects the divergence in the density of states, though
$F(\Omega)$ remains finite around $\Omega\simeq \pm B$ and $F(\pm
B)=0$. 
It can be seen in Fig.\ref{Fig5}(b) that $F_0(\Omega)$ is the 
dominant contribution to $F(\Omega)$ for
$\Omega\simeq E_0$ and we see that the antisymmetric spectral shape of
the interference component of $F_0(\Omega)$ (green line)
is the origin of the asymmetry in $F(\Omega)$.
In fact, neglecting $\mbox{Im}[N_d]$ in Eqs.(\ref{F0}), which is
appropriate except for $|E_0| > B$, the interference term in $F_0(\Omega)$ is approximately
given by
\begin{eqnarray}\label{F0inter}
    &&F_0^{inter}(\Omega)
    =\frac{2T_{dc}T_{0c}}{gB}
    \mbox{Re}[N_d]\gamma  \nonumber\\
    &&\times \left\{ 
    \frac{\Omega-\tilde{E}_0}{(\Omega-\tilde{E}_0)^2+\gamma^2} + 
    \frac{\tilde{E}_0-E_0}{(\Omega-\tilde{E}_0)^2+\gamma^2} 
    \right\}
    \; .
\end{eqnarray}
When the condition $1-|E_0|/B \gg g$ is satisfied, 
it holds that $\mbox{Re}[N_d] \simeq 1$
and $\tilde{E}_0-E_0 \ll 1$.
As a result, only the first term of 
Eq.(\ref{F0inter}) contributes to $F_0^{inter}$ (Fig.\ref{Fig5}(b), green line),
leading to the antisymmetric spectral shape of the
interference component of $F_0(\Omega)$.
The asymmetric spectral shape obtained when the branch point effect is neglected 
is represented by the well-known Beutler-Fano profile~\cite{Fano61}. 

However when $E_0$ lies close to the band edges, $|\tilde{E}_0-E_0|$
becomes large, the second term of Eq.(\ref{F0inter}b) can no longer be
neglected, which increases the interference component of $F_0(\Omega)$
compared to the $d$-$d$ diagonal component.
We show in Fig.\ref{Fig6} the calculated results for the same
parameters as in Fig.\ref{Fig4}(b) where $E_0=-0.98B$, except that we take
$T_{dc}=1.0$.
As seen in Fig.\ref{Fig6}(b) the interference component of $F_0(\Omega)$  (green
line) is enhanced compared with that in 
Fig.\ref{Fig5}(b).
As mentioned above, in this case quantum coherence plays a key role in the
decay process, which is clearly reflected through the enhancement of the
interference effect in the absorption spectrum.

\section{Summary and discussions}

In the present work, we have evaluated the charge transfer rate from an 
adatom impurity on a 1D semiconductor superlattice. 
The decay rate is dramatically enhanced due to two square-root forms of
singularities in the density of states.  In the vicinity of the singularities
at either edge of the band spectrum, the decay rate becomes a nonanalytic 
function of the coupling constant $g$ at $g=0$ as it can only be 
expanded in powers of $g^{4/3}$.
The time evolution of the localized adatom state is governed by the
pole and branch point contributions to the decay rate, which account for
the exponential and nonexponential decay, respectively.

We have demonstrated that the absorption spectrum from the inner core
level of the adatom to the conduction states is an appropriate probe
to observe both contributions.
While the branch point effect is usually small, it becomes important 
when the adatom localized state is resonant with the miniband of the
1D superlattice in the vicinity of the singularity in the density of states.
In the case where the branch point effect is significant, the quantum
coherence in the decay process becomes exaggerated.
This explains 
the enhancement of the quantum interference in the absorption spectrum.

The advantage of using the semiconductor superlattice is that it is 
easy to vary the parameters in an electronic system.
Using modern nanotechnology, we can vary the parameter values 
widely, in order to systematically investigate the effect
on the decay process.
Other spectroscopic techniques, such as the resonant optical light
scattering spectrum,  may also be used to investigate the time
evolution of the unstable state in detail, which we are now studying.

\acknowledgements

We  thank Professor V. Kocharovsky, Dr. G. Ordonez, Professor W.
Schieve and Professor G. Sudarshan for insightful discussions.
We acknowledge  the
Engineering Research Program of the Office of Basic Energy Sciences
at the U.S. Department of Energy, Grant No DE-FG03-94ER14465 for
supporting  this work.
This work was supported by the Grant-in-Aid for Scientific Research from
the Ministry of Education, Science, Sports, and Culture of
Japan.


\clearpage

\begin{figure}
\begin{center}
\caption{(a) An adatom attached to a 1D quantum-dot array, and (b) level
structures of the adatom localized state and the bound state in each 
quantum well. The width of each well is a few nm to 100
nm. The adatom  is located at the $n=0$-th
well. 
The two optical absorption transitions from the core level $|c\rangle$ with $T_{dc}$ and 
$T_{0c}$ are also shown by the arrows.} 
\label{Fig1} \end{center}
\end{figure}

\begin{figure}
\begin{center}
\caption{(a) $E_i/B$   and $\tilde{E}_0 /B$ vs.  $E_0/B$ for $g=0.5$, which are
shown by the thick solid and dashed lines, respectively. 
The location of the critical values $\pm E_\gamma/B$ are indicated by
the arrows. 
The thin line is $y=E_0/B$. 
(b) $\gamma/B$ vs. $E_0/B$ for $g=0.1$.     
The maximum value of  $\gamma_{max}/B$  occurs at $E_0/B=\pm 1$.} 
\label{Fig2} \end{center}
\end{figure}

\begin{figure}[tbp]
    \centering
    \caption{The contours of the integral for $F(\Omega)$
    Eq.(\ref{FOmega2}) (a) and its deformation (b), and the integral for 
    $F_0(\Omega)$ Eq.(\ref{F0}a) (c). }
    \label{Fig3}
\end{figure}

\begin{figure}[tbp]
    \centering
    \caption{The calculated $F(\Omega)$ (solid line), $F_0(\Omega)$
    (dashed line), and $F_1(\Omega)$ (chain line) for $g=0.2$, and $T_{dc}=1.0$ and
    $T_{0c}=0$: (a) $E_0=-0.1B$ and
    (b) $E_0=-0.98B$. The thin vertical lines indicate the position of
    $E_0$. In (b), the horizontal axis is expanded around
    $\Omega\simeq -B$, while the overall spectrum is shown in the inset.
}
    \label{Fig4}
\end{figure}

\begin{figure}[tbp]
    \centering
    \caption{The calculated $F(\Omega)$ (a), $F_0(\Omega)$
    (b), and $F_1(c)$ for the same parameters for Fig.\ref{Fig5}(a) except $T_{0c}=1.0$: 
    The spectra are decomposed into the $d$-$d$ diagonal (red line), $\tilde{0}$-$\tilde{0}$
    diagonal (blue line), and the interference terms (green line).}
    \label{Fig5}
\end{figure}

\begin{figure}[tbp]
    \centering
    \caption{The parameters here are the same as in Fig.\ref{Fig5}(b) except that $T_{0c}=1.0$.}
    \label{Fig6}
\end{figure}

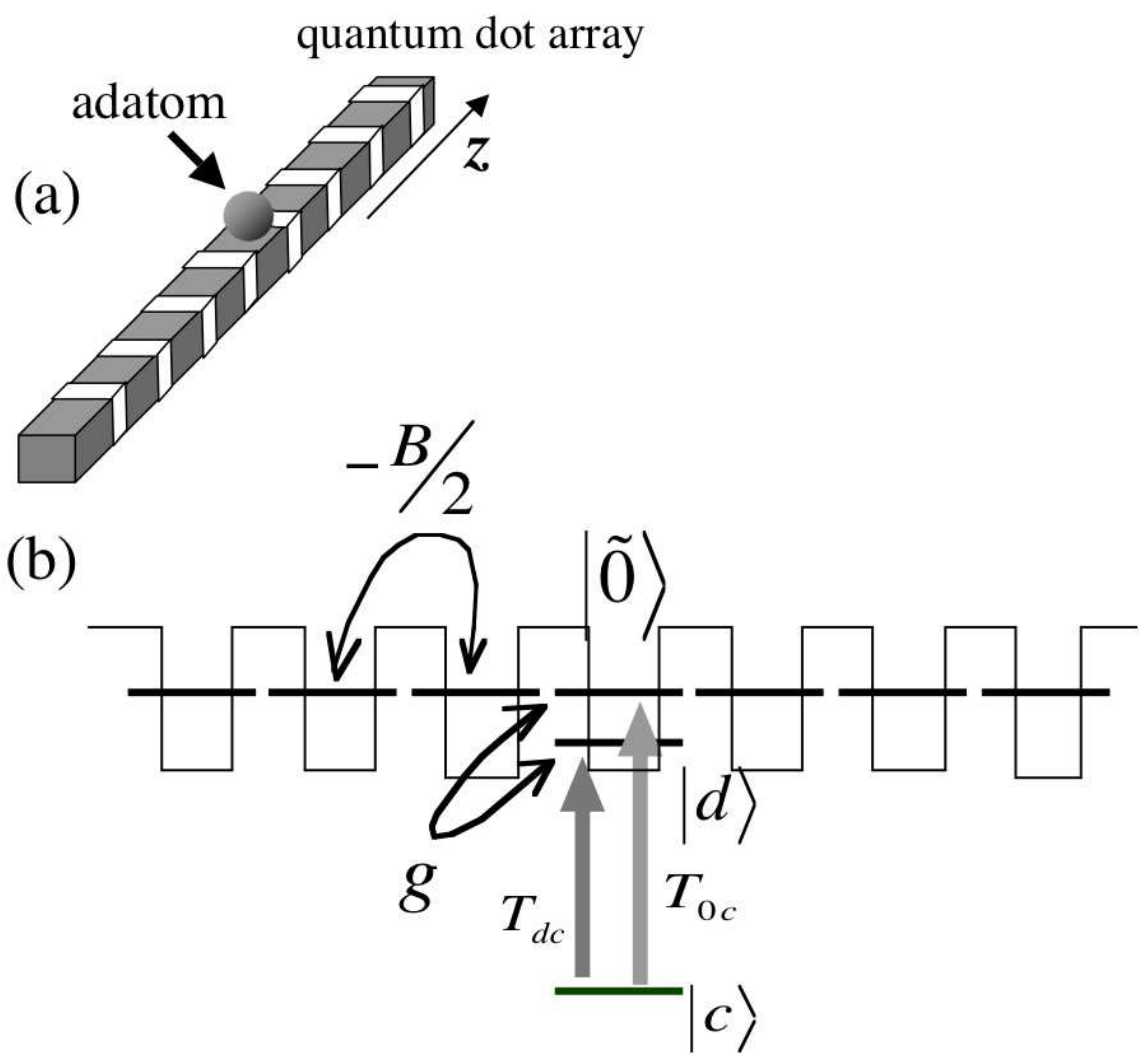
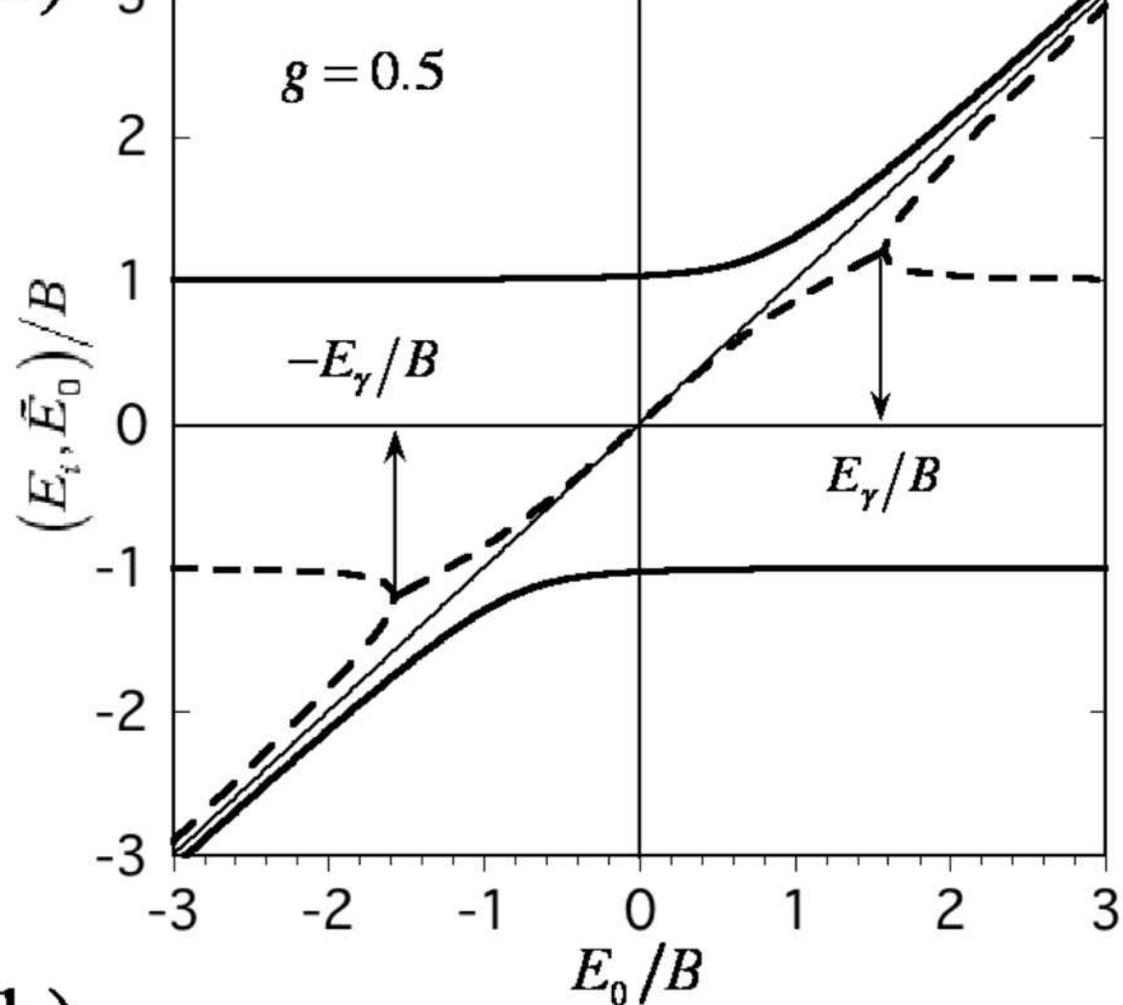
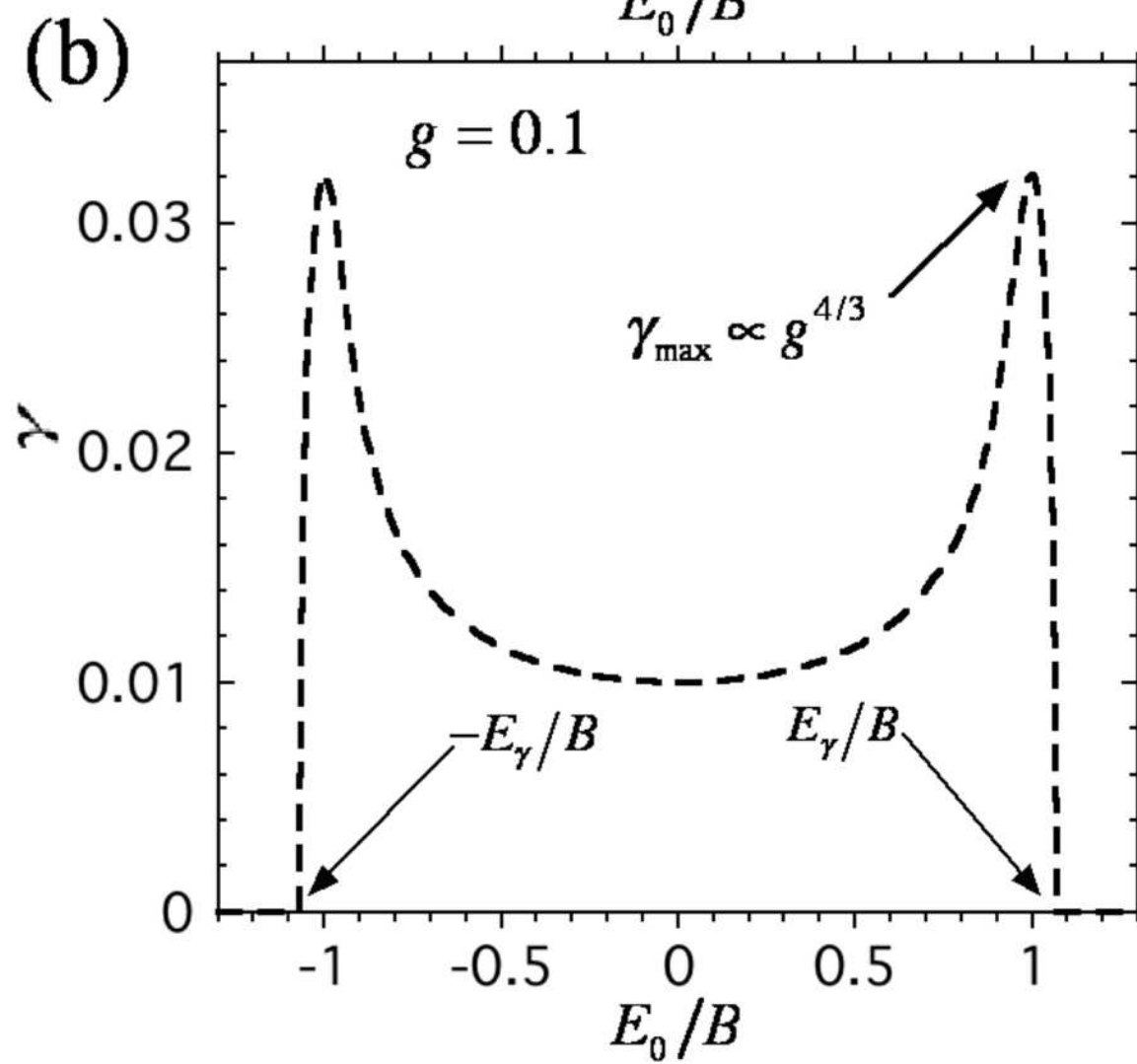
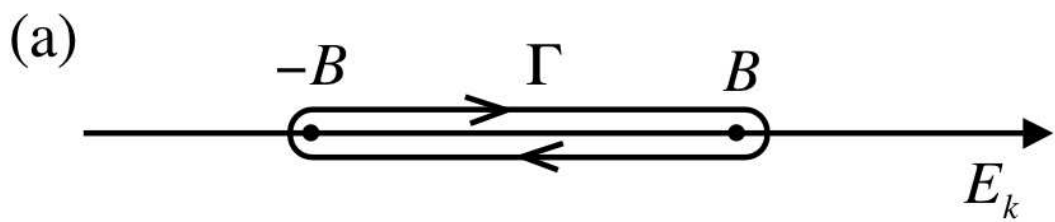
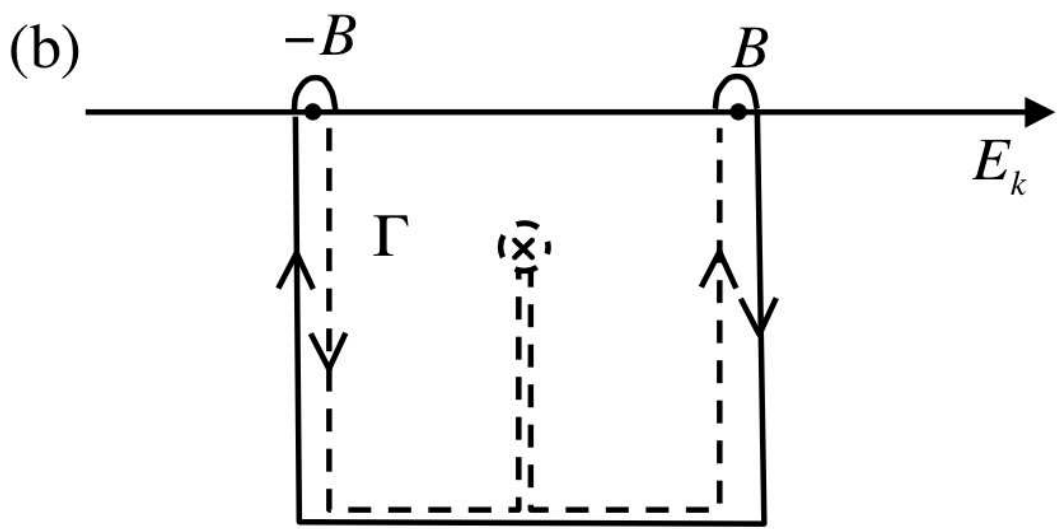
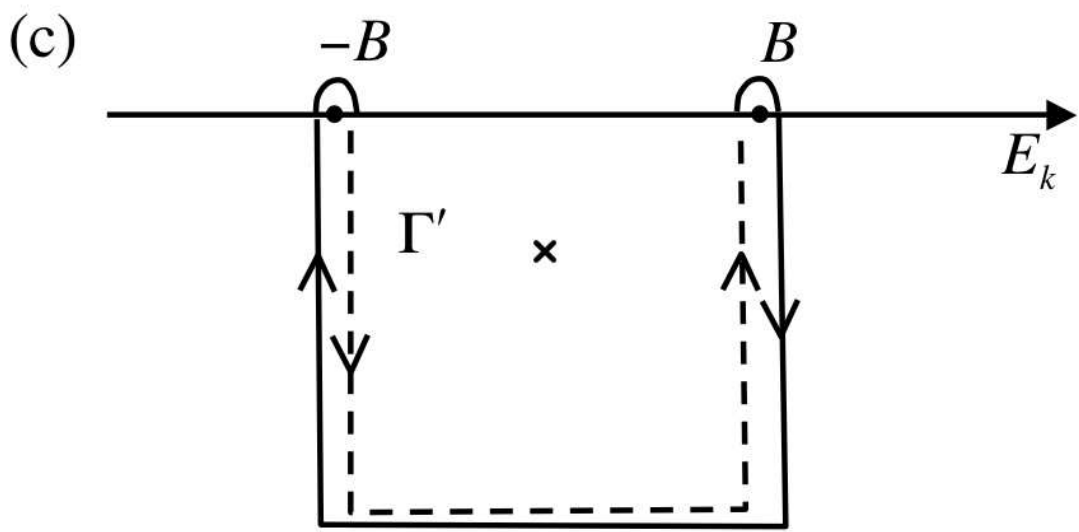
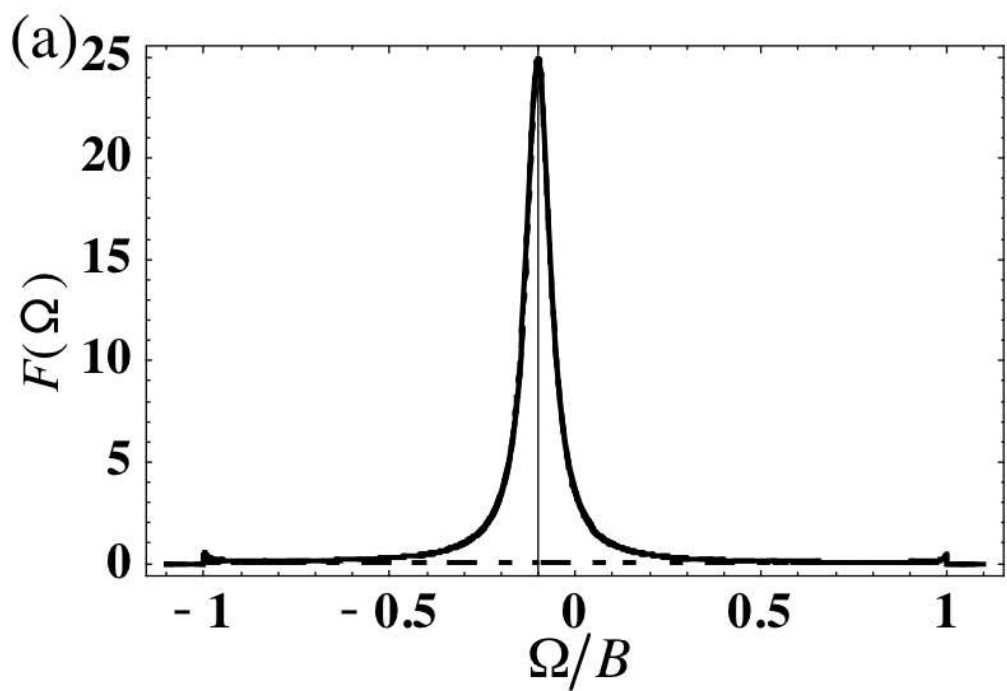
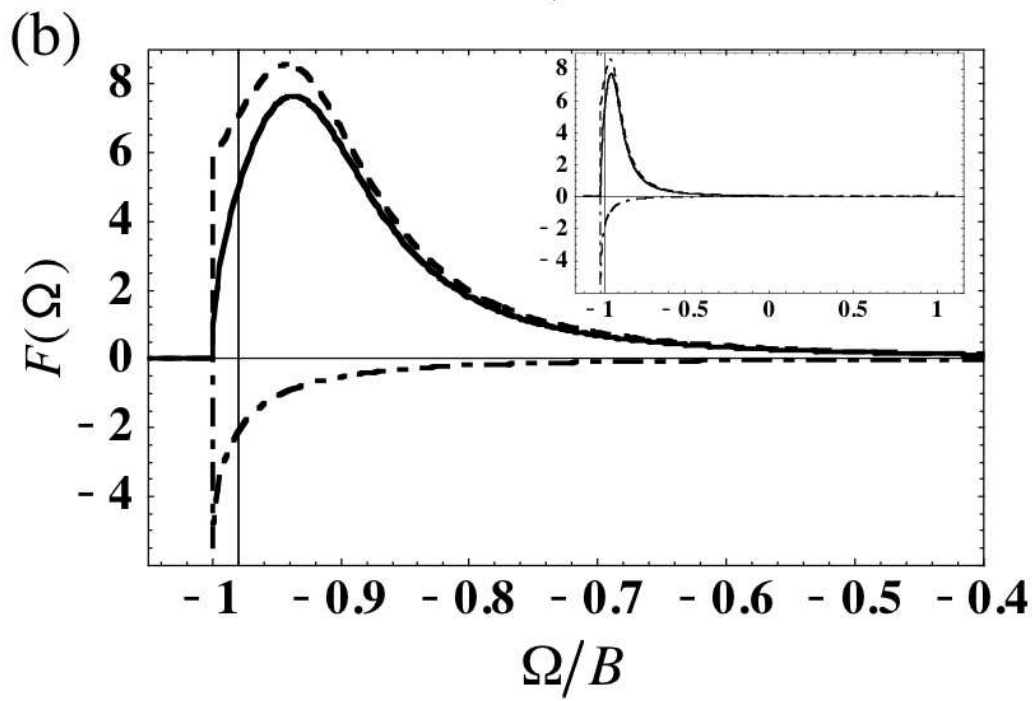
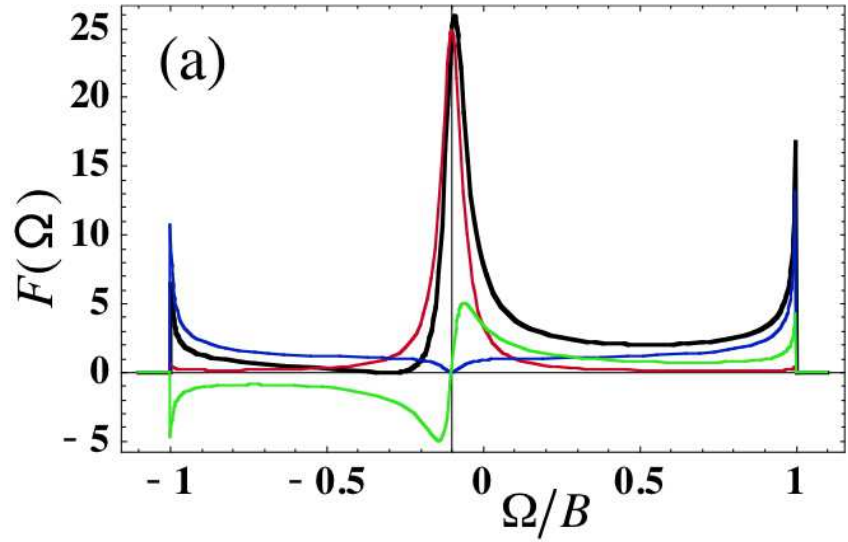
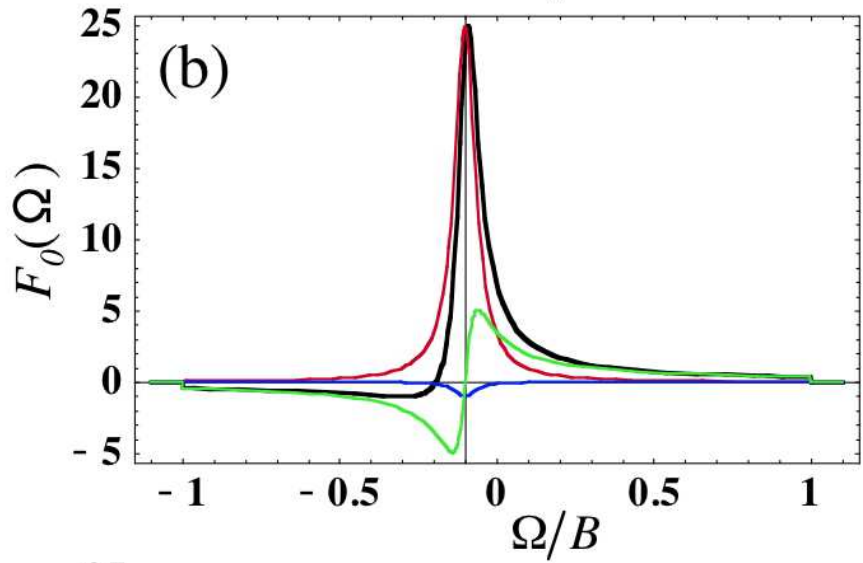
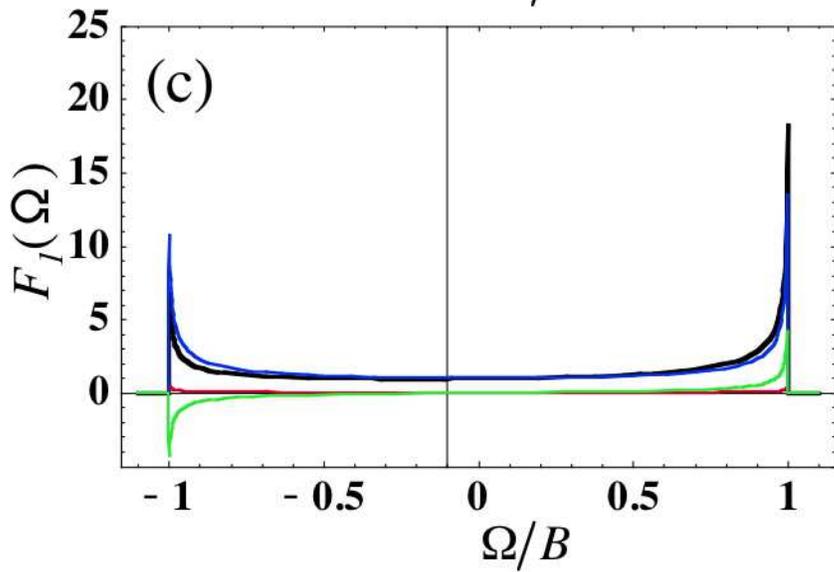
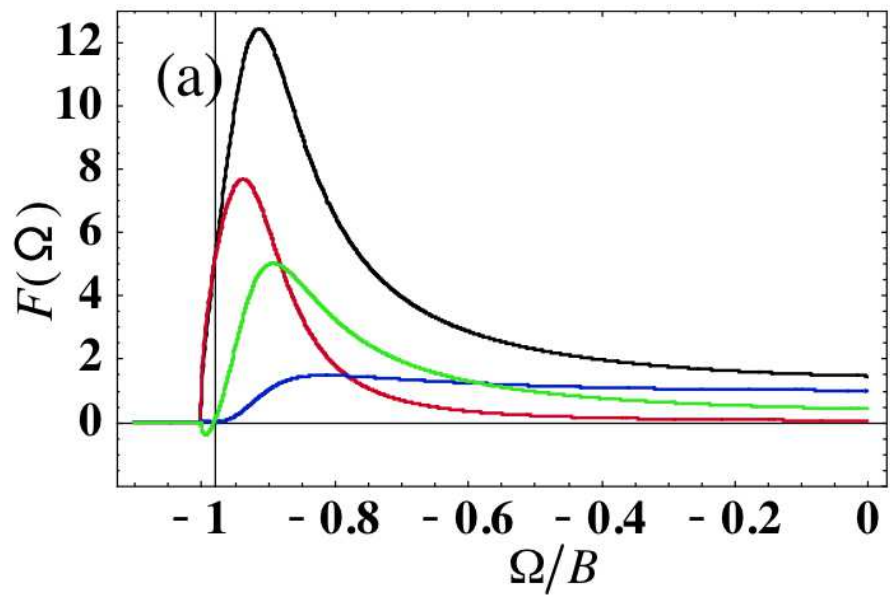
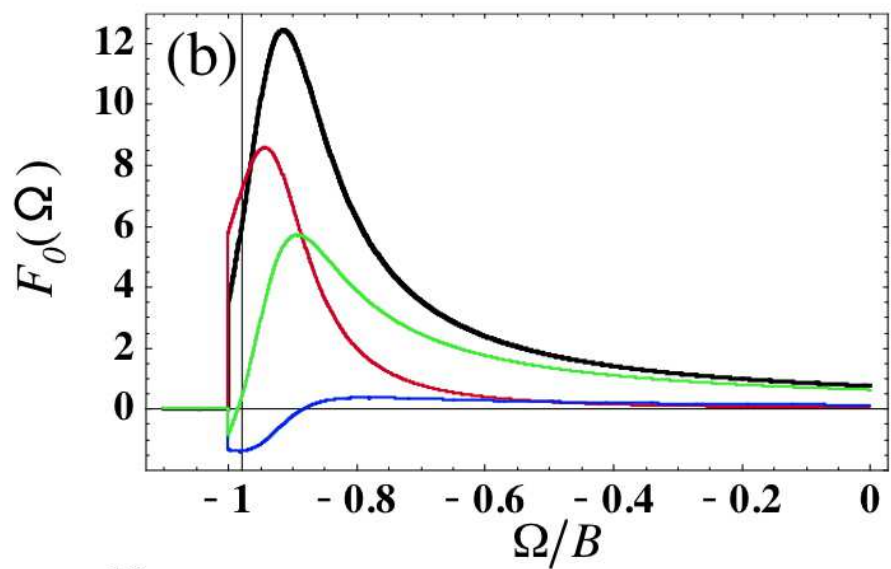
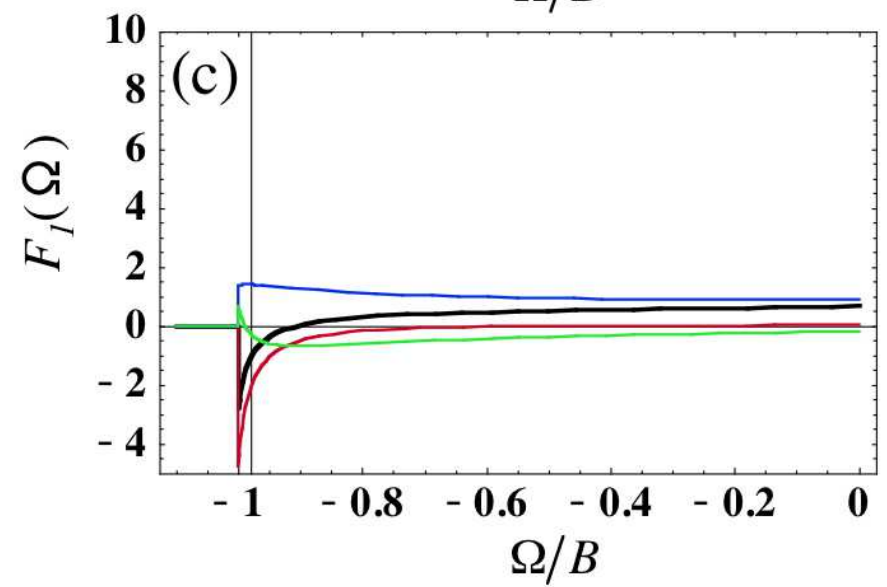
\end{document}